\documentclass[vecphys]{svmult}

\usepackage{makeidx}         
\usepackage{graphicx}        
\usepackage{multicol}        
\usepackage[bottom]{footmisc}
\usepackage{amssymb}

\makeindex             

\begin{document}

\title*{Near-UV Merger Signatures in Early-Type Galaxies}
\titlerunning{NUV Merger Signs in ETGs}
\author{Jodie R. Martin\inst{1}
\and Robert W. O'Connell\inst{2}
\and J. E. Hibbard\inst{3}}
\institute{University of Virginia Charlottesville, VA  \texttt{jodie@virginia.edu}
\and University of Virginia Charlottesville, VA  \texttt{rwo@virginia.edu}
\and National Radio Astronomy Observatory Charlottesville, VA  \texttt{jhibbard@nrao.edu}}
%
%
\maketitle

\begin{abstract}

Hierarchical assembly of early-type galaxies (Es and S0s) over an extended 
period of time will result in mixed-generation stellar populations.  Here we 
look for signatures of composite populations in broad-band, near-ultraviolet 
(2500-3400 \AA), high-resolution HST imaging of the cores of 12 bright 
early-type galaxies without obvious dust or active galactic nuclei.  Near-UV 
imaging is a sensitive probe for the detection of younger components with ages 
in the range of 10 Myr to 5 Gyr.  Only two galaxies have central colors 
($r<\,0.75\,r_e$) that are consistent with a single-generation population.  
The other ten require a composite population.

\end{abstract}

\section{Introduction}
\label{sec:1}





%

In $\Lambda$CDM cosmology galaxies are assembled hierarchically over an 
extended period by mergers of smaller systems.  All galaxies, including early 
type galaxies (ETGs), are expected to contain multi-epoch stellar 
populations.  Toomre \& Toomre \cite{17} were the first to suggest that some 
early types could form from the interactions of disk galaxies.  They 
predicted that a dynamical redistribution of stars could transform merging 
disk galaxies into an elliptical galaxy, though such a redistribution of a 
collisionless system would not increase the maximum density to that observed 
in elliptical cores \cite{9}.  It was later realized that the gaseous 
components of the progenitor galaxies would collide, compress, and rapidly 
flow into the center of the gravitational potential well \cite{6,7,16}, where 
they can form new stars, thereby increasing the central density of the 
remnant.  

There is considerable evidence supporting a merger origin for at least some 
fraction of the ETG population.  Observations of the disk-disk merger remnant 
NGC7252 show a surface brightness profile as well as a maximum density similar 
to an elliptical galaxy, not to a disk galaxy \cite{14}.  Molecular gas 
densities in three merger remnants similarly indicate evolution toward an ETG 
remnant \cite{4}.  A study of the K-band light profiles of coalesced merger 
remnants found excess stellar light \cite{12} in the centers of some of the 
galaxies. Monochromatic photometric surveys of nearby ETGs \cite{1,10,11} have 
serendipitously discovered that 20-40\% have excess light above a smooth 
profile in the center-most regions, and many nearby ETGs show dynamical 
disturbances that correlate with color \cite{15}. Finally, at higher redshifts 
ETGs with blue cores \cite{5} and the E+A galaxies \cite{2,8} are likely 
post-merger systems.  

If these central light excesses result from mergers, they are likely to 
contain intermediate-age stellar populations.  In this project, we are using 
the Hubble Space Telescope (HST) to study that possibility by obtaining 
near-UV light profiles and colors of two groups of ETGs: a main sample with 
excess light in their centers and a control sample without.  We emphasize the 
near-UV band because it is the most sensitive to intermediate-age stellar 
populations.

\section{Sample Selection and Observations}
\label{sec:2}

Twelve galaxies were selected from published HST imaging studies of ETGs 
\cite{1,10,11}.  We chose six galaxies with light profiles that rise smoothly 
with decreasing radius and six with apparent excess light at small radii above 
a smooth profile. All twelve were chosen to have no known AGNs and no 
discernible central dust. The sample includes seven E's, one dE, and four S0 
galaxies.  

The twelve galaxies were observed in Cycle 13 with HST using the ACS-HRC in 
three filters: F250W (``NUV''), F330W (``U''), and F555W (``V'').  The 
exposure times were selected to achieve similar sensitivity levels across all 
bands.  Each exposure was dithered in four positions per filter for sub-pixel 
sampling and cosmic ray cleaning.  The pixel scale of the HRC is 
$0.027^{\prime\prime}$/pix with a $27^{\prime\prime}\times27^{\prime\prime}$ 
field of view and the full-width half maximum of the point spread function 
(PSF) is 0.0459$^{\prime\prime}$ (1.7 pixels) for all three filters.  Features 
larger than 2.5-6.1 parsecs in the sample galaxies are resolved. 

\section{Data Processing and Stellar Population Models}
\label{sec:3}

The images were initially processed and combined through the HST pipeline 
routines CALACS and Multidrizzle.  Isophotes were fit to the V images and then 
applied to the U and NUV images to extract their light profiles; derived 
colors are therefore for annular elliptical apertures.  Sky background 
estimates were made from two sources - an average over blank field exposures 
extracted from the HST archive, and an average over the outer $5\times10^4$ 
pixels of each galaxy image.  The minimum of the two estimates was used.  The 
observed images in each band were smoothed with the other two filters' PSFs, 
since each PSF differs slightly in its substructure.  

In this contribution, we discuss only the integrated colors of the galaxies.  
The total light within a radius of 0.75 $r_e$ of the smoothed images was 
measured in each band, where $r_e$ is the effective radius measured in the 
B-band\cite{18}.  The colors were corrected for foreground reddening.  The 
effect of the anomalous central component on integrated colors was checked by 
masking out the central 0.015 $r_e$ and remeasuring the fluxes.  The 
deviations were less than 0.013 mag in both the NUV-V and U-V colors.  

We use the P\'egase stellar population synthesis models \cite{3} to interpret 
the galaxy colors.  Spectral energy distributions (SEDs) are available for a 
pre-determined grid of ages, $\tau$, and metallicities, Z, in the form of 
monochromatic luminosities per unit mass, [{\it S}\,]=(erg/s/\AA/$M_\odot$).  
These were computed for an instantaneous burst with a Salpeter initial mass 
function \cite{13} over the mass range of $0.1-120\,M_\odot$.  We select the 
subset of models that covers the age range of 10 Myr to 15 Gyr and the 3 
highest metallicities, Z=[0.4,1,2.5]$\; Z_\odot$.  The synthetic SEDs were 
integrated over the HST bandpasses to get total normalized luminosities for 
each age and metallicity.

To interpret our photometry, we compare the locations of the galaxy colors to 
the P\'egase single-generation models and also to simple dual-generation 
models made up of an old population of age $\tau_{O}$ and a young population 
of age $\tau_{Y}$ that contributes a fraction of the total mass of the galaxy 
$f_{Y}$.  We assume that the old and young populations have the same 
metallicity.  Our grid of mixed models is computed as follows:

\begin{center}
$S_{Mix}(Z,\tau_{Y},f_{Y},\lambda)=f_{Y}\times S(Z,\tau_{Y},\lambda)+
(1-f_{Y})\times S(Z,\tau_{O},\lambda)$
\end{center}

\noindent where $f_{Y}$ evenly covers the log fraction space between $10^{-4}$ 
to $9\times10^{-1}\;$.  Most of the models assume $\tau_{O}= 12$ Gyr.  Note 
that a single-generation stellar population corresponds to the case $f_{Y}=1$.

\section{Results}
\label{sec:4}

Figures 1 and 2 compare the integrated galaxy colors to the single- and 
dual-generation stellar population models.  In Figure 1, the single-generation 
models are plotted as open symbols connected by solid lines and the galaxies 
as solid diamonds and circles.  The plot provides a consistency check on the 
photometry and the models in the sense that no objects lie significantly below 
or to the right of the single-generation model locus; this region is 
``forbidden'' to composite populations.  The locus corresponding to simple 
mixed-metallicity models --- i.e.\ populations of a small age range (say 10-12 
Gyr) but a large range of Z --- is narrow and closely follows the 
single-generation line. 

In Figure 2, the single-generation populations are shown again along with 
tracks of dual-generation populations for a 12 Gyr old population and five 
younger components (10 Myr [solid lines], 100 Myr, 400 Myr, 1 Gyr, 4 Gyr 
[dashed lines]).  The locus for the 10 Myr component represents an upper 
envelope to the region that is consistent with a dual-generation model for a 
given old population age.  The sensitivity of the NUV to small amounts of 
light from young populations is indicated by the large volume of color-color 
space enclosed within the upper envelopes.  In general, many dual-generation 
models are consistent with a given data point in this region, but limits can 
be placed on their components.

Only two galaxies, NGC 3377 and NGC 3384, have colors which are consistent 
with single-generation populations, and these require $Z\gtrsim Z_\odot$.  All 
of the other galaxies require mixed-generation populations.  Eight of these 
require a dominant population with $Z > Z_\odot$, assuming no components can 
be older than 12 Gyr.  The corresponding limit for NGC 4482 and NGC 4239 is $Z
\gtrsim 0.5Z_\odot$.  To exemplify the stellar population mixture 
interpretations, we find 3 galaxies (NGCs 2778, 4478, and 4570) are well 
described by old, supersolar populations mixed with young components of ages 
250, 500, and 450 Myr, and total mass contributions of 0.4, 1.5, and 1.2$\%$, 
respectively.

Our basic result is that most of the galaxies show strong evidence of a 
composite population with contributions from components much younger than 
10-12 Gyr.  This is expected if most early type galaxies have formed by 
mergers over extended periods.  Interestingly, this is true whether or not the 
galaxies possess excess central light suggestive of a late merger event 
(filled circles versus diamonds in Figs. 1 \& 2).  Although the three largest 
departures from the single generation locus occur for excess light systems 
(NGCs 4239, 4482, and 7457), the UV/optical color properties of the other 
members of the two groups are similar.  The colors also show that, because the 
excess central light feature does not affect the total integrated colors, the 
new stars formed during a merger event are far more efficiently mixed 
throughout the merger remnant than the numerical models predict \cite{6,16}.

%
%
%
%
%
%
%
%
%
%
%
%
%
\begin{figure}
\centering
\includegraphics{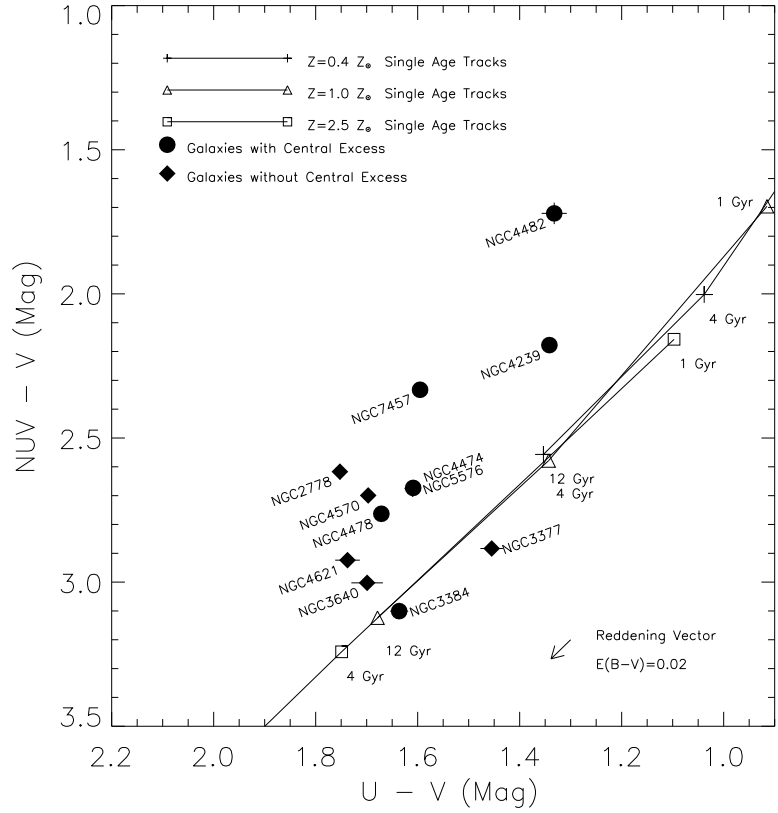}
\caption{Color-Color plot comparing the total integrated colors of the 12 
galaxies to the single-generation single-metallicity P\'egase models and the 
tracks of single-generation mixed-metallicity models for a few select ages, 
[1 Gyr, 4 Gyr, 12 Gyr].}
\label{fig:1}       
\end{figure}
\begin{figure}
\centering
\includegraphics{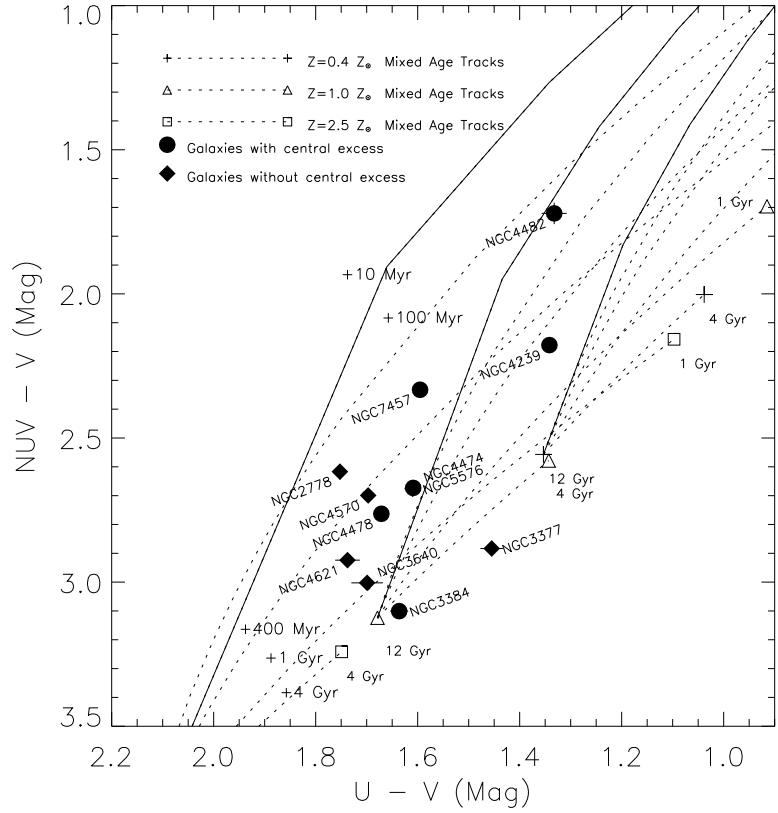}
\caption{Color-Color plot comparing the total integrated colors of the 12 
galaxies to the single-generation single-metallicity P\'egase models and the 
tracks of dual-generation single-metallicity models for a few select ages, 
[+100 Myr, +400 Myr, +1 Gyr, +4 Gyr].  For each metallicity, the track for +10 
Myr is shown as a solid line for emphasis.}
\label{fig:2}       
\end{figure}


\printindex
\end{document}